\documentclass{emulateapj}

\begin{document}

\shorttitle{The Initial-Final Mass Relation}
\shortauthors{Kalirai et al.}

\title{The Initial-Final Mass Relationship: Spectroscopy of White Dwarfs in NGC 2099 
(M37)\altaffilmark{1}}

\author{Jasonjot Singh Kalirai\altaffilmark{2}, Harvey B. Richer\altaffilmark{2}, David Reitzel\altaffilmark{3}, Brad~M.~S.~Hansen\altaffilmark{3,4}, R. Michael Rich\altaffilmark{3}, Gregory~G.~Fahlman\altaffilmark{5}, 
Brad~K.~Gibson\altaffilmark{6}, \& Ted von Hippel\altaffilmark{7}}

\notetoeditor{If accepted, we would appreciate that this Letter and the following Letter 
entitled ``The Dearth of Massive, Helium-rich White Dwarfs in Young Open Star Clusters'' 
are published back to back, with this Letter coming first.}

\altaffiltext{1} {Based on observations with Gemini (run ID GN-2002B-Q-11) and Keck.
Gemini is an international partnership managed by the Association of
Universities for Research in Astronomy under a cooperative agreement with
the National Science Foundation.  The
W. M. Keck Observatory, which is operated as a scientific partnership
among the California Institute of Technology, the University of
California, and NASA, was made possible by the generous financial
support of the W. M. Keck Foundation.}
\altaffiltext{2}{Department of Physics and Astronomy, 6224 Agricultural Road, University of British 
Columbia, Vancouver, BC, V6T 1Z4, Canada; jkalirai@astro.ubc.ca, richer@astro.ubc.ca.}
\altaffiltext{3}{Department of Physics and Astronomy, University of California at Los Angeles, 
Box 951547, Knudsen Hall, Los Angeles, CA 90095-1547; reitzel@astro.ucla.edu, hansen@astro.ucla.edu, 
rmr@astro.ucla.edu.}
\altaffiltext{4}{Alfred P. Sloan Research Fellow.}
\altaffiltext{5}{National Research Council of Canada, Herzberg Institute of Astrophysics, 5071 West
Saanich Road, RR5, Victoria, BC V9E 2E7, Canada; greg.fahlman@nrc.gc.ca.}
\altaffiltext{6}{Centre for Astrophysics and Supercomputing, Swinburne University, P.O. Box 218, 
Hawthorn, VIC 3122, Australia; bgibson@swin.edu.au.}
\altaffiltext{7}{Department of Astronomy, University of Texas at Austin, RLM 15.308, C-1400, Austin 
TX 78712-1083; ted@astro.as.utexas.edu.}
 
 \slugcomment{\it 
}

\lefthead{Kalirai, J. S. et al.}
\righthead{Initial-Final Mass Relation}

\begin{abstract}
We present new observations of very faint white dwarfs (WDs) in the rich open star 
cluster NGC 2099 (M37).  Following deep, wide field imaging of the cluster 
using {\sl CFHT}, we have now obtained spectroscopic observations of candidate 
WDs using both {\sc gmos} \normalfont on {\sl Gemini} and {\sc lris} 
\normalfont on {\sl Keck}.  Of our 24 WD candidates (all fainter than 
$V$ = 22.4), 21 are spectroscopically confirmed to be bona fide WDs, 4-5 
of which are most likely field objects.  Fitting 18 of the 21 WD spectra with model 
atmospheres, we find that most WDs in this cluster are quite massive (0.7--0.9 $M_\odot$), 
as expected given the cluster's young age (650 Myr), and hence, high turn-off mass 
($\sim$ 2.4 $M_\odot$).  We determine a new initial-final mass relationship and almost 
double the number of existing data points from previous studies.  The results indicate that 
stars with initial masses between 2.8 and 3.4 $M_\odot$ lose 70--75\% of their mass through stellar evolution.  
For the first time, we find some evidence of a metallicity dependence on the 
initial-final mass relationship.

\end{abstract}


\section{Introduction}

The initial-final mass relationship connects the mass of the final products of 
stellar evolution for intermediate mass stars, white dwarfs (WDs), to their progenitor mass.  
It is a required input for the determination of the ages and distances of globular clusters 
from modeling their WD cooling sequences (Hansen et al.~2004), for constraining 
chemical evolution in galaxies, for determining supernova rates (van den Bergh \& 
Tammann 1991), and for understanding feedback processes and star formation in galaxies 
(e.g., Somerville \& Primack 1999). Yet, despite its fundamental importance,
this relation remains poorly constrained observationally.

The first attempt to map the relation was made by Weidemann (1977) by comparing theoretical 
models (e.g., Fusi-Pecci \& Renzini 1976) of mass loss to the masses of a 
few WDs in both the Pleiades and Hyades star clusters. Since then most of the
work has focused on using 
observations of WDs in young open star clusters to provide empirical
constraints on the relationship (Koester \& Reimers 1981, 1985, 1993, 1996;
Reimers \& Koester 1982, 1989, 1994; Weidemann \& Koester 1983; Weidemann 1987,
Jeffries 1997). 
This effort, spanning almost two 
decades, is summarized in Weidemann (2000).  The result is a monotonically 
increasing relationship based on about 20 data points, from observations 
of about a dozen star clusters.  

A few recent studies, such as Claver et al. (2001) and Williams et al. (2004), have 
been successful in finding 
a half dozen WDs each in the Praesepe and NGC 2168 clusters.  The 
synthesis of these data with the earlier studies results in an initial-final 
mass relationship displaying a fair amount of scatter.  For example, depending 
on whose data one uses, initial mass stars of 3--4 $M_\odot$ can produce WDs 
ranging anywhere from 0.65--0.8 $M_\odot$.  This result suggests that the 
relationship may have a stochastic component.

It is desirable to find a young, rich star cluster with a large number of WDs 
that can be spectroscopically studied.  NGC 2099 is such a cluster, with 50 
WD cluster candidates measured using imaging observations (Kalirai et al. 
2001a).  The cluster is very rich, containing over 4000 stars, and has a main-sequence 
turn-off of $\sim$ 2.4~$M_\odot$.  Its distance modulus is $(m {\rm-}M)_{V}$ = 11.5, and 
its reddening is $E(B {\rm-}V)$ = 0.23.  In this {\it Letter}, we present a new WD initial-final 
mass relationship based on 18 WDs in NGC 2099.  In the following {\it Letter}, we address 
the surprising result that all WDs in NGC 2099 (as well as those found 
in other young open clusters) are all DA spectral type (Kalirai et al. 2005).

\section{Observations} \label{obswdsspec}

Imaging and spectroscopic observations of NGC 2099 were obtained with the
Canada-France-Hawaii ({\sl CFH}), {\sl Gemini North}, and {\sl Keck I}
telescopes.  In our wide field {\sl CFHT} imaging study (Kalirai et al.~2001a),
we found $\sim$~67 WD candidates in the central 15$'$ of NGC 2099.  Based on 
comparisons between the cluster field and a blank field surrounding the cluster, 
we estimate the rate of field star contamination among our candidates to be 
$\sim$~25\% (thus yielding 50 cluster WD candidates).  We then plotted the locations of all of 
these faint-blue objects on the sky and obtained further observations of
3 smaller sub-fields in the cluster.  These fields were chosen to maximize the
number of WD candidates (all objects in the faint-blue end of the color-magnitude 
diagram [CMD] were treated 
as WD candidates).  With {\sl Gemini}, we imaged three 5${.}^{\prime}$5 $\times$
5${.}^{\prime}$5 fields using the {\sc gmos} \normalfont multi-object imaging/spectroscopic
instrument (Murowinski et al.~2003).  With {\sl Keck}, we imaged the same three fields 
with the {\sc lris} \normalfont imaging/spectroscopic instrument that has a 
5$' \times$ 7$'$ field of view (Oke et al. 1995).  These imaging data were not 
significantly deeper than the original {\sl CFHT} data and were only used to 
ensure astrometric accuracy for the spectroscopy.
 
Multi-object spectra were obtained for a single {\sl Gemini} field, and for two of 
the {\sl Keck} fields.  The {\sl Gemini} observations used the B600 grating, which 
simultaneously covers 2760 (centered at $\sim$ 4700 ${\rm \AA}$).  
The data were binned by a factor of four in the spectral direction to improve the 
signal-to-noise ratio (S/N).   The {\sl Keck} observations use the 600/4000 grism 
(blue side) which simultaneously covers 2580 ${\rm \AA}$ (centered at 4590 ${\rm 
\AA}$).  For the {\sl Gemini} field, we obtained 22 individual 1-hour exposures spread 
over 22 days, taken mostly at low air-masses ($<1.2$) and good seeing 
($\sim$ 0${.}^{\prime\prime}$8).  For the {\sl Keck} fields, we obtained 4 
2000 s exposures in each of the two fields, also at sub-arcsecond seeing.

The {\sl Gemini} spectroscopic data were reduced using the {\sl Gemini} {\sc iraf}
\normalfont package, version 1.3.  The {\sl Keck} data were reduced using a set 
of python routines written by D. Kelson (2004, private communication; 
Kelson 2003).  The individual exposures were bias subtracted, flat-fielded, cleaned for cosmic rays, 
wavelength calibrated, sky subtracted, extracted, combined, and flux calibrated 
(using bright standard stars) within each of these programs.  Details of the steps 
involved in each of these procedures will be provided in a forthcoming paper (J. S. Kalirai 
et al. 2005, in preparation).  The only major problem occurred for some of the {\sl Keck} field 2 
spectra, which were taken at high air-masses and so the bluest flux was lost as a 
result of atmospheric dispersion.  Fortunately, two of the three stars that we were fitting 
from this field have also been observed in the higher S/N {\sl Gemini} data.

\begin{figure}
\begin{center}
\leavevmode
\includegraphics[width=7cm]{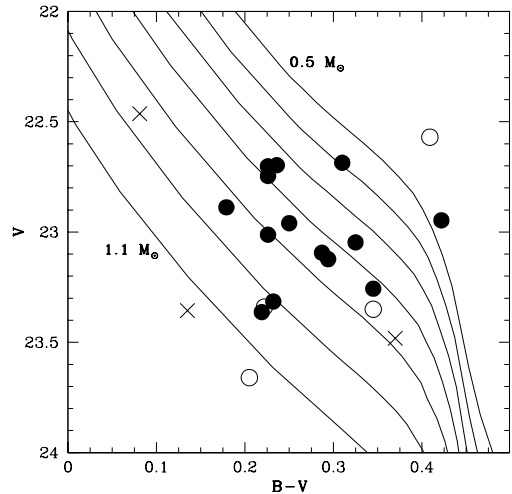}
\end{center}
\caption{Faint-blue end of the CMD of NGC 2099 shown with all spectroscopically 
confirmed WDs in this work (21 objects - based on {\sl CFHT} photometry).  The 14 filled 
circles are those with consistent theoretical and observed magnitudes (assuming 
cluster membership), the 4 open circles have inconsistent theoretical 
magnitudes, and the 3 crosses are objects that we could not fit to spectral 
models.  The model sequences represent WD cooling models (Fontaine, Brassard, 
\& Bergeron 2001) at increments of 0.1 $M_\odot$, with 0.5 $M_\odot$ at the top and 
1.1 $M_\odot$ at the bottom.}
\label{cmd}
\end{figure}

In total, we obtained spectroscopy of 24 individual WD candidates in the field
of NGC 2099 (3 of these stars turned out not to be WDs).  Therefore, despite sampling 
only 14\% of the total cluster area, we include almost 1/3 of the total WD population 
(cluster and field) given the careful positioning of the fields.  This is therefore
the largest individual star cluster WD sample that has ever been spectroscopically
acquired.  A CMD showing the locations of the NGC 2099 WDs 
is shown in Figure \ref{cmd} (based on {\sl CFHT} data).  The 18 circles represent those 
objects that we were able to spectroscopically fit (see next section).  Four of 
these, shown as open circles, represent those objects that have inconsistent 
theoretical magnitudes (from fitting the spectra) as compared to the observed 
magnitudes, assuming they are cluster members.  Three WDs, which we could not 
fit with models, are shown as crosses.  All of the complete spectra are shown 
in the companion {\it Letter} (Kalirai et al. 2005).  

\section{Analysis} \label{analysis}

\begin{figure}
\begin{center}
\leavevmode
\includegraphics[width=8cm]{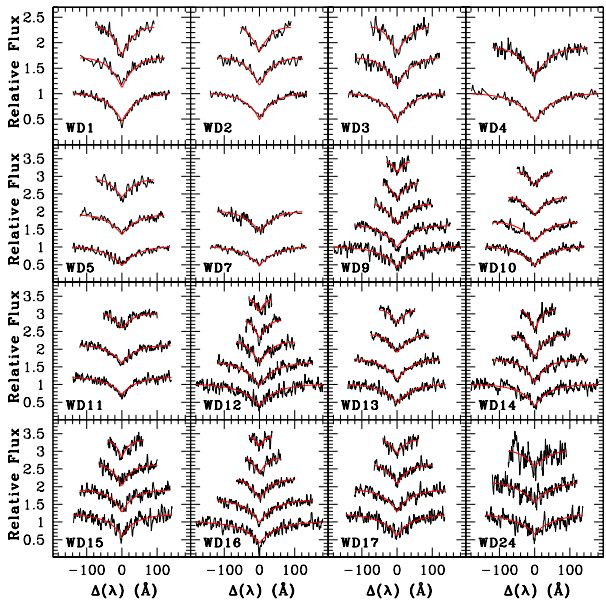}
\end{center}
\caption{Spectroscopic fits (red) shown for the hydrogen Balmer lines of 16 WDs in NGC 2099.  
Within each panel, the lines are $H_\beta$ (bottom), $H_\gamma$, $H_\delta$, $H_\epsilon$, and 
$H_8$ (top).  The number of higher order lines varies depending on the spectral coverage 
of each star.  The first six objects (WD1 -- WD7) have the highest S/Ns and were 
obtained using {\sl Gemini}.  The remaining stars ({\sl Keck} objects) have lower 
S/Ns but have bluer spectral coverage (hence the fitting of higher order 
lines).  The identifications are the same as those in Kalirai et al. (2005), in which 
we present the full spectrum for all WDs.  The corresponding $T_{\rm eff}$, log~$g$, and 
$M_{\rm f}$ of these WDs are given in Table 1.  The masses of two additional objects are measured 
as discussed in the text.}
\label{spectrafits}
\end{figure}

Using the techniques described in Bergeron, Saffer, \& Liebert (1992), we determine 
$T_{\rm eff}$ and log~$g$ for each WD.  The line profiles 
are first normalized using two points on the continuum on either side of each 
absorption line.  Therefore, the fit should not be affected by the flux calibration 
unless there is a strange ``kink'' or slope change at the location of a Balmer line.  
The fitting of the line shapes uses the nonlinear least-squares method of 
Levenberg-Marquardt (Press et al.~1986).  The $\chi^{2}$ statistic is calculated 
and minimized for combinations of $T_{\rm eff}$ and log~$g$, using normalized 
model line profiles of all absorption lines simultaneously.  The resulting 1-$\sigma$ 
errors in $T_{\rm eff}$ and log~$g$ were tested by simulating synthetic spectra with 
the same number of absorption lines, and similar S/Ns, and measuring the 
output parameters from fitting these spectra.  These results are found to have 
errors slightly less than those in the true spectra (as expected given the small errors in 
flux calibration and other defects), and so we use the true spectra errors.  Masses ($M_{\rm f}$) 
and WD cooling ages ($t_{\rm cool}$) are found by using the updated evolutionary models of Fontaine, 
Brassard, \& Bergeron (2001) for thick hydrogen layers ($q(\rm H)$ = $M_{\rm H}/M$ = 
10$^{-4}$) and helium layers of $q(\rm He)$ = 10$^{-2}$.  The core is assumed to be 
a 50/50 C/O mix.  In Figure \ref{spectrafits} we present the model atmosphere fits for 
each WD.  

The WD cooling age represents the time that each of these stars has spent traversing 
from the tip of the asymptotic giant branch (AGB) down to its present WD luminosity.  We 
can now calculate the progenitor main-sequence lifetime ($t_{\rm ms}$, the total lifetime 
of the star up to the tip of the AGB) assuming an age for the cluster.  In Kalirai et al. (2001a), 
we fitted the NGC 2099 main-sequence to solar metallicity isochrones and determined an 
age of 520 Myr.  Recently, C. Deliyannis et al.~(2004, private communication) have 
spectroscopically measured the cluster metallicity to be subsolar ($Z$ = 0.011 $\pm$ 0.001), 
and the reddening to be $E(B {\rm-}V)$ = 0.23 $\pm$ 0.01, slightly larger than the 
value that we used.  Using these parameters, the age of NGC 2099 is now calculated to be 
650 Myr using the same Ventura et al.~(1998) models and procedure as described in Kalirai et 
al.~(2001a).  We also derive similar ages using the Padova group (620 Myr - Girardi et al.~2000) 
and Yale-Yonsei isochrones (630 Myr - Yi et al.~2001) for this metallicity (J. S. Kalirai et al. 
2005, in preparation).

The $t_{\rm ms}$ for each star is determined by subtracting the WD cooling ages 
($t_{\rm cool}$) from the cluster age (650 Myr).  The initial progenitor masses 
($M_{\rm i}$) for the WDs are then calculated using the Hurley, Pols, \& Tout~(2000) 
models for $Z$ = 0.01.  For this metallicity, these models give very similar values to those 
derived from the Ventura et al.~(1998) models.  For these masses, a 100 Myr cluster age 
difference would only cause a $\sim$0.3 $M_\odot$ initial mass 
difference (for 350 Myr lifetimes).  This is a very small effect on our initial-final mass 
relationship, and therefore the results are not highly sensitive to the derived cluster age.  
Table 1 summarizes the derived parameters for each star, with 1-$\sigma$ error bars.  Also 
given are the theoretical magnitudes (from fitting the spectra) and the observed 
magnitudes [assuming $(m {\rm-}M)_{V}$ = 11.5 to NGC 2099].
 
\begin{deluxetable*}{lllllllllllll}
\tablecolumns{12} 
\tablewidth{0pc} 
\tablecaption{Derived Parameters of WDs \label{BigTab1}}
\tablehead{\colhead{Star} & \colhead{$\alpha$} & \colhead{$\delta$} & \colhead{$T_{\rm eff}$} & \colhead{log ($g$)} & \colhead{$M_{\rm f}$} & \colhead{$M_{V}$} & \colhead{$M_{V}$} & \colhead{$t_{\rm cool}$} & \colhead{$t_{\rm ms}$} & \colhead{$M_{\rm i}$} \\ \colhead{} & \colhead{(J2000)} & \colhead{(J2000)} & \colhead{(K)} & \colhead{} & \colhead{($M_\odot$)} & \colhead{(theory)} & \colhead{(obs.)} & \colhead{(Myr)} & \colhead{(Myr)} & \colhead{($M_\odot$)}}

\startdata 

WD1 & 05:52:26.46 &  32:30:47.9 & 16500 $\pm$ 800 & 7.75 $\pm$ 0.19 & 0.48 $\pm$ 0.09 & 10.72 & 11.07 & 111 $\pm$ 32 & 539 $\pm$ 32 & $2.79^{+0.05}_{-0.06}$ \\

WD2 & 05:52:17.16 &  32:29:04.1 & 19900 $\pm$ 900 & 8.11 $\pm$ 0.16 & 0.69 $\pm$ 0.10 & 10.90 & 11.20 & 108 $\pm$ 36 & 542 $\pm$ 36 & $2.76^{+0.08}_{-0.05}$ \\

WD3 & 05:52:36.25 &  32:32:51.5 & 18300 $\pm$ 900 & 8.23 $\pm$ 0.21 & 0.76 $\pm$ 0.13 & 11.21 & 11.39 & 179 $\pm$ 64 & 471 $\pm$ 64 & $2.92^{+0.16}_{-0.14}$ \\

WD4 & 05:52:34.53 &  32:29:12.4 & 16900 $\pm$ 1100 & 8.40 $\pm$ 0.26 & 0.87 $\pm$ 0.15 & 11.65 & 11.51 & 329 $\pm$ 133 & 321 $\pm$ 133 & $3.36^{+0.76}_{-0.40}$ \\

WD5 & 05:52:25.11 &  32:30:55.7 &  18300 $\pm$ 1000 & 8.33 $\pm$ 0.22 & 0.83 $\pm$ 0.14 & 11.39 & 11.62 & 224 $\pm$ 85 &  426 $\pm$ 85  &  $3.02^{+0.26}_{-0.19}$ \\

WD6$^*$ & 05:52:18.08 & 32:29:38.4 & 17200 $\pm$ 1600 & \nodata & 0.92 $\pm$ 0.15 & 11.76 & 11.76 & 300 $\pm$ 100 & 350 $\pm$ 100 & $3.25^{+0.44}_{-0.28}$ \\

WD7 & 05:52:36.96 &  32:33:29.9 & 17800 $\pm$ 1400 & 8.42 $\pm$ 0.32 & 0.88 $\pm$ 0.19 & 11.58 & 11.82 & 303 $\pm$ 148 & 347 $\pm$ 148  &  $3.26^{+0.78}_{-0.40}$ \\

WD9 & 05:52:25.14 & 32:40:03.6 & 15300 $\pm$ 400 & 8.00 $\pm$ 0.08 & 0.61 $\pm$ 0.05 & 11.20 & 11.19 & 202 $\pm$ 28 & 448 $\pm$ 28 & $2.97^{+0.06}_{-0.07}$ \\

WD10 & 05:52:25.82 & 32:36:03.8 & 19300 $\pm$ 400 & 8.20 $\pm$ 0.07 & 0.74 $\pm$ 0.04 & 11.08 & 11.20 & 131 $\pm$ 19 &  519 $\pm$ 19 &  $2.81^{+0.04}_{-0.02}$ \\

WD11 & 05:52:26.57 & 32:35:25.4 &  23000 $\pm$ 600 & 8.54 $\pm$ 0.10 & 0.96 $\pm$ 0.06 & 11.34 & 11.25 & 153 $\pm$ 30 & 498 $\pm$ 30 & $2.86^{+0.07}_{-0.07}$  \\

WD12 & 05:52:11.96 & 32:40:39.5 & 13300 $\pm$ 1000 & 7.91 $\pm$ 0.12 & 0.55 $\pm$ 0.07 & 11.34 & 11.45 & 313 $\pm$ 70 & 337 $\pm$ 70 & $3.30^{+0.30}_{-0.22}$ \\

WD13 & 05:52:16.21 & 32:38:18.5 &   18200 $\pm$ 400 & 8.27 $\pm$ 0.08 & 0.79 $\pm$ 0.05 & 11.30 & 11.46 & 183 $\pm$ 28 & 467 $\pm$ 28 & $2.93^{+0.06}_{-0.07}$\\

WD14 & 05:52:06.43  & 32:38:29.4 & 11400 $\pm$ 200 & 7.73 $\pm$ 0.16 & 0.45 $\pm$ 0.08 & 11.38 & 11.55 & 318 $\pm$ 61 & 332 $\pm$ 61 & $3.31^{+0.26}_{-0.20}$ \\

WD15$^{**}$ & 05:52:27.42  & 32:35:50.0 & 11300 $\pm$ 200 & 8.35 $\pm$ 0.15 & 0.83 $\pm$ 0.09 & 12.30 & 11.85 & 786 $\pm$ 189 & \nodata $\pm$ \nodata & $>3.73$ \\

WD16 &  05:52:22.89  & 32:36:29.4 & 13100 $\pm$ 500 & 8.34 $\pm$ 0.10 & 0.83 $\pm$ 0.06 & 11.97 & 11.86 & 542 $\pm$ 94 & 108 $\pm$ 94 &  $5.20^{+\ ?}_{-1.20}$ \\

WD17$^{**}$ & 05:52:29.29  & 32:37:06.7 & 12300 $\pm$ 400 & 8.72 $\pm$ 0.10 & 1.05 $\pm$ 0.05 & 12.77 & 12.16 & 1346 $\pm$ 256 & \nodata $\pm$ \nodata & $>7$ \\

WD21$^*$ & 05:52:26.42 & 32:30:04.3 & 17200 $\pm$ 2000 & \nodata & 0.85 $\pm$ 0.15 & 11.59 & 11.59 & 261 $\pm$ 100 & 389 $\pm$ 100 & $3.13^{+0.36}_{-0.25}$ \\

WD24 & 05:52:25.34 & 32:29:35.2 & 18700 $\pm$ 1100 & 8.82 $\pm$ 0.24 & 1.11 $\pm$ 0.13 & 12.24 & 11.84 & 492 $\pm$ 237 & 158 $\pm$ 237 & $4.43^{+\ ?}_{-1.32}$ \\

\enddata 

\tablenotetext{*}{Masses for these two WDs are determined by assuming they are cluster members - see \S \ref{ifmr}.}
\tablenotetext{**}{The cooling time for these two WDs ($t_{\rm cool}$) is larger than the cluster age - see \S \ref{ifmr}.}

\end{deluxetable*} 

\newpage

\section{The Initial-Final Mass Relationship} \label{ifmr}

\begin{figure}
\begin{center}
\leavevmode
\includegraphics[width=7cm]{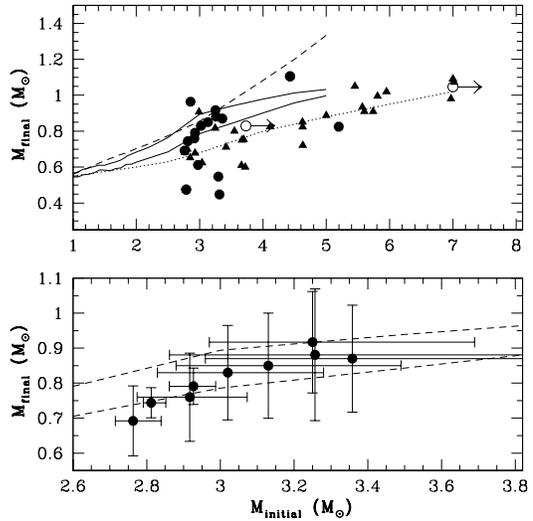}
\end{center}
\caption{Top - WD initial-final mass relationship shown for the 18 
WDs spectroscopically fitted in this work (circles) and all previous constraints 
(triangles).  Also shown are several semi empirical and theoretical relations 
as discussed in the text.  Bottom - Closer look at those WDs that form the tight 
sequence with M$_{\rm i}$ = 2.8--3.4 $M_\odot$ and are bound by the two Marigo~(2001) 
theoretical relations.}
\label{fig.ifmr}
\end{figure}

In Figure \ref{fig.ifmr} (top) we present the initial-final mass relationship for 
the sixteen WDs in Figure \ref{spectrafits} (filled circles), as well as two other 
objects, WD6 and WD21 (see Kalirai et al. 2005 for the spectra of these 
objects).  For these two stars, we could not determine an accurate log~$g$ as a result 
of the spectra having too low S/Ns.  However, the effective temperatures are well 
constrained, and therefore we derive the masses of the stars by combining this 
information with the luminosities (assuming they are cluster members).  This gives 
the radius, which coupled with a mass-radius relation, 
gives the mass of the stars.  The masses of these two stars, as well as their 
progenitor masses, are found to be in good agreement with others in the cluster 
(WD6 has $M_{\rm f}$ = 0.92 and $M_{\rm i}$ = 3.25, and WD21 has $M_{\rm f}$ = 0.85 
and $M_{\rm i}$ = 3.13).  Two of the WDs shown in Table 1 have negative progenitor 
lifetimes.  This is due to the WD cooling age of these stars being larger than the 
cluster age.  For WD17, we artificially set its initial mass to 7 $M_\odot$ 
(the most massive star that produces a WD in the Ventura et al.~1998 models).  
For WD15, we compute a 95\% confidence lower limit of 408 Myr for $t_{\rm cool}$ 
and determine the initial mass based on this cooling age.  These stars are both 
plotted with open circles and an arrow pointing to higher masses to reflect 
the lower limits.  It is unlikely that these objects are field WDs given 
their unusually high masses (see Madej, Nalezyty, \& Althaus 2004 for field 
WD mass distribution).

Figure \ref{fig.ifmr} (top) shows that, with just one star cluster, we have nearly doubled 
the number of data points on the initial-final mass plane.  This is remarkable 
considering the time and effort required to establish the previous constraints (shown 
as triangles).   Furthermore, half of our data points sit along a very tight 
sequence with initial masses in the range 2.8--3.4 $M_\odot$ and final masses in the 
range 0.7--0.9 $M_\odot$.  Fitting these data points and their two dimensional errors to 
a straight line, we determine a slope of 0.33 and an intercept of -0.19.  The residual 
rms of the fit is 0.03.  This suggests that stars with masses between 2.8 and 
3.4~$M_\odot$, and metallicities slightly less than solar, will lose 70--75\% 
of their mass through stellar evolution.

Four of our stars are seen well below the tight sequence of points discussed above.  Although 
only one of these four stars has an inconsistent theoretical magnitude, they could still 
be field WDs located near the cluster.  Not only are their masses consistent 
with the distribution of field WDs ($M$ = 0.56 $M_\odot$ - Sloan Digital Sky Survey data; 
Madej, Nalezyty, \& Althaus 2004), but we also expect 4-5 WDs in our sample to be field 
stars based on WD number counts in blank fields surrounding NGC 2099 (Kalirai et al. 2001a).  
The lower masses of these stars could also be attributed to binary star evolution and 
mass transfer.  We will investigate these possibilities further in J. S. Kalirai et al. 
(2005, in preparation).

We have plotted several theoretical and semi empirical relations in Figure 
\ref{fig.ifmr}.  The dotted line is the revised semi empirical initial-final 
mass relationship from Weidemann (2000).  The dashed line is the theoretical 
initial-final mass relationship from the Padova group stellar evolution 
models (Girardi et al.~2000; L. Girardi 2004, private communication) for $Z$ = 
0.008.  The two solid lines are theoretical initial-final mass relations from 
Marigo (2001), for $Z$ = 0.008 (top) and $Z$ = 0.02 (bottom).  The latter models 
have improved mass-loss mechanisms in their post-main sequence evolutionary phases and 
therefore should be preferred over the others (L. Girardi 2004, private 
communication).  Considering this, it is interesting to note that half of 
our NGC 2099 cluster WDs reside in a region of the initial-final mass 
relationship bound by the two Marigo (2001) relations (see bottom panel of Figure \ref{fig.ifmr}).  
This is what we would expect, given that the cluster metallicity falls between that used in 
the two models ($Z$ = 0.011).  We point out, however, that the errors in our measurements 
are too large for us to be able to discriminate between the two relations.  
Our data points are, however, mostly located above the other previous constraints.  
This can also be explained by the fact that almost all of the other clusters are solar or higher 
metallicity, resulting in more efficient mass loss during stellar evolution (Marigo 2001).  
Although this could be a systematic effect (e.g., the masses of the WDs in other clusters have 
been derived using different models), these results may be suggesting that, for the first 
time, we are seeing the effects of metallicity on the initial-final mass relationship.  
In J. S. Kalirai et al. (2005, in preparation), we will eliminate these systematics by 
reevaluating all of the previous data on the initial-final mass relation (including the 
new results of Williams, Bolte, \& Koester 2004) using a consistent fitting method and set 
of evolutionary models.  

The spectroscopic measurement of masses of WDs with $V \simeq$ 23 has not been previously 
accomplished.  The number of targets that are accessible at these magnitudes is several orders 
of magnitudes larger than previously identified (see work on the on-going {\sl CFHT} Open 
Star Cluster Survey; Kalirai et al. 2001a; 2001b; 2001c; 2003).  With continuing observations 
of rich open star clusters, we can envision placing $>$100 data points into this very fundamental 
relation.  By observing both younger and older clusters, the entire initial progenitor mass 
range can be constrained and a detailed initial-final mass relation can be produced. 

\newpage

\acknowledgements
We would like to thank Pierre Bergeron for providing us with his
models and spectral fitting routines.  This project could not have
succeeded without this input.  We also wish to thank Dan Kelson, Inger
Jorgenson, and Jarrod Hurley for providing help with various parts of this
project.  J.~S.~K. received financial support during this work through an
NSERC PGS-B research grant.  The research of H.~B.~R. is supported in part
by NSERC.  B.~M.~S.~H. is supported by NASA grant ATP03-0000-0084.  B.~K.~G.
acknowledges the financial support of the Australian Research Council.
T.~v.~H. appreciatively acknowledges support from NASA through LTSA grant
NAG5-13070.


\begin{references}

\reference{} Bergeron, P., Saffer, R. A., \& Liebert, J. 1992, ApJ, 394, 228 

\reference{} Claver, C. F. et al.~2001, ApJ, 563, 987

\reference{} Fontaine, G., Brassard, P., \& Bergeron, P. 2001, PASP, 113, 409

\reference{} Fusi-Pecci, F. \& Renzini, A. 1976, A\&A 46, 447

\reference{} Girardi, L. et al.~2000, A\&AS, 141, 371

\reference{} Hansen et al.~2004, accepted in ApJS, astro-ph/0401443

\reference{} Hurley, J. R., Pols, O. R., \& Tout, C. A. 2000, MNRAS, 315, 543

\reference{} Jeffries, R. D. 1997, MNRAS, 288, 585

\reference{} Kalirai, J. S. et al. 2005, submitted to ApJL

\reference{} Kalirai, J. S. et al. 2003, AJ, 126, 1402

\reference{} Kalirai, J. S. et al. 2001a, AJ, 122, 3239

\reference{} Kalirai, J. S. et al. 2001b, AJ, 122, 266

\reference{} Kalirai, J. S. et al. 2001c, AJ, 122, 257

\reference{} Kelson, D. 2003, PASP, 115, 808, 688

\reference{} Koester, D. \& Reimers, D. 1996, A\&A, 313, 810

\reference{} Koester, D. \& Reimers, D. 1993, A\&A, 275, 479

\reference{} Koester, D. \& Reimers, D. 1985, A\&A, 153, 260    

\reference{} Koester, D. \& Reimers, D. 1981, A\&A, 99, L8

\reference{} Madej, J., Nalezyty, M., \& Althaus, L. G. 2004, A\&A, 419, L5 

\reference{} Marigo, P. 2001, A\&A, 370, 194

\reference{} Murowinski, R. et al. 2003, 2003, SPIE, 4841, 1440

\reference{} Oke, J. B. 1995, PASP, 107, 375

\reference{} Press, W. H. et al.~1986, Numerical Recipes, Cambridge: Cambridge Univ. Press

\reference{} Reimers, D. \& Koester, D. 1994, A\&A, 285, 451

\reference{} Reimers, D. \& Koester, D. 1989, A\&A, 218, 118

\reference{} Reimers, D. \& Koester, D. 1982, A\&A, 116, 2, 341

\reference{} Somerville, R. S. \& Primack, J. R. 1999, MNRAS, 310, 1087

\reference{} van den Bergh, S. \& Tammann, G. 1991, ARA\&A, 29, 363

\reference{} Ventura, P. et al.~1998, A\&A, 334, 953

\reference{} Weidemann, V. 2000, A\&A, 363, 647

\reference{} Weidemann, V. 1987, A\&A, 188, 74

\reference{} Weidemann, V. \& Koester, D. 1983, A\&A, 121, 77

\reference{} Weidemann, V. 1977, A\&A, 59, 411

\reference{} Williams, K. A., Bolte, M., \& Koester, D. 2004, ApJ, 615, L49

\reference{} Yi, S. et al.~2001, ApJS, 136, 417


\end{references}
\end{document}